\documentclass[a4paper,fleqn]{cas-sc}

\def\tsc#1{\csdef{#1}{\textsc{\lowercase{#1}}\xspace}}
\tsc{WGM}
\tsc{QE}
\tsc{EP}
\tsc{PMS}
\tsc{BEC}
\tsc{DE}

\begin{document}
\let\WriteBookmarks\relax
\def\floatpagepagefraction{1}
\def\textpagefraction{.001}
\shorttitle{Pontus-Mpemba effect in cavity  quantum electrodynamics}
\shortauthors{S. Longhi}

\title [mode = title]{Pontus-Mpemba effect in cavity  quantum electrodynamics}                      
\author[1,2]{Stefano Longhi}[type=,
                        auid=000,bioid=1,
                        prefix=,
                        role=,
                        orcid=0000-0002-8739-3542]
\ead{stefano.longhi@polimi.it}


\affiliation[1]{organization={Dipartimento di Fisica, Politecnico di Milano},
                addressline={Piazza Leonardo da Vinci 32}, 
                city={Milano},
                postcode={20133}, 
                state={},
                country={Italy}}

\affiliation[2]{organization={Institute for Cross-Disciplinary Physics and Complex Systems, University of the Balearic Islands and the Spanish National Research Council},
                addressline={Cra. de Valldemossa, km 7.5}, 
                city={Palma},
                postcode={07122 }, 
                state={Illes Balears},
                country={Spain}}

\begin{abstract}
The quantum Pontus-Mpemba effect is a counterintuitive phenomenon in which a quantum system relaxes faster through a two-step evolution protocol than through a single, unquenched relaxation. This work proposes its realization in cavity quantum electrodynamics using the Jaynes-Cummings model with photon loss. The model captures the coherent interaction between a two-level atom and a single quantized mode of a lossy cavity, providing a minimal yet realistic setting to explore dissipative quantum dynamics. Restricting the analysis to the single-excitation sector, the dynamics feature damped vacuum Rabi oscillations for weak dissipation that transition to near-exponential atomic decay under strong dissipation. A sudden quench of the cavity decay rate generates distinct relaxation trajectories from the same initial atom-cavity state. The atomic excitation then displays a non-monotonic, accelerated decay, where a trajectory with a quenched dissipation relaxes faster than fixed-loss evolution. The effect originates from the interplay between coherent atom-photon exchange and cavity dissipation, establishing a clear and experimentally accessible realization of the quantum Pontus-Mpemba effect in both optical and circuit QED platforms.
\end{abstract}




\maketitle

\section{Introduction}

Non-equilibrium systems exhibit a wealth of counterintuitive phenomena that challenge conventional notions of relaxation and equilibration. Among the most remarkable is the \emph{Mpemba effect} \cite{r1,r2,r3,r4}, originally observed in classical systems, in which a system initially farther from equilibrium relaxes faster than one closer to equilibrium. This phenomenon has been documented in diverse classical settings \cite{r4,r5,r6,r7,r8,r9,r9b,r9c,r10,r11,r12}, including supercooled liquids, granular fluids, clathrate hydrates, and nanoscale resonators, and has inspired a large body of theoretical and experimental work \cite{r4,r12}. Although various mechanisms have been proposed -- such as overlap with slow-decaying modes, accelerated relaxation pathways, and thermal overshooting -- a universally accepted explanation remains elusive. The concept of the \emph{quantum Mpemba effect} (QME) generalizes this idea to quantum systems \cite{r12b}, where anomalous relaxation can occur either in open quantum systems, coupled to an environment \cite{
r13,r14,r15,r16,r17,r19,r20,r21,Referee1,Referee2,r28,r29,r30,r31,r34,r35,r36,r37,r38,r40,r42,r43,r44,r45,r45b,r50,r51,r52,r52b,r53,r54,arxiv:2512.02709}, or in closed quantum systems, evolving under unitary dynamics \cite{r18,
r18,r23,r24,r25,r32,r33,r26,r27,r39,r41,r46,r47,r48,r49}. In both cases, the effect is typically associated with distinct initial states, with more ``excited'' or ``asymmetric'' initial states exhibiting faster relaxation. The QME has been theoretically investigated in spin chains, many-body localized systems, non-Hermitian systems, and random quantum circuits, and has been experimentally observed in several quantum platforms, confirming its relevance beyond theoretical models \cite{r12b}.

Recently, a further generalization, the \emph{Pontus-Mpemba effect} (PME), has been proposed \cite{P1,P2,P3,P4,P5}. Unlike standard Mpemba protocols, the PME occurs for \emph{identical initial states}, with the counterintuitive acceleration emerging from a \emph{two-step evolution protocol}. In this scheme, one system relaxes directly toward the target state under a given relaxation (Liouvillian) operator, while the second system first evolves under a quenched or modified operator toward an intermediate state, and is subsequently driven to the same target state. The PME is realized when the total relaxation time of the two-step protocol is shorter than that of the direct single-step relaxation. This protocol offers several conceptual advantages over standard Mpemba scenarios, including the absence of a ``parameter distance'' requirement and the ability to optimize intermediate states to enhance the effect.

In this work, we propose a realization of the quantum Pontus-Mpemba effect in a physically relevant quantum-optical system, namely the dissipative Jaynes--Cummings model \cite{JC1,JC2,JC3,JC4} describing a two-level atom coupled to a single cavity mode with photon loss. The dynamics are restricted to the single-excitation sector, where coherent atom--photon exchange gives rise to damped vacuum Rabi oscillations for weak dissipation or near-exponential atom decay at strong dissipation (Purcell effect). By implementing a sudden quench of the cavity decay rate, from high to low cavity $Q$ factor, two relaxation trajectories are generated starting from the same initial atom--cavity state. It is shown that, in appropriate parameter regimes, the two-step protocol accelerates the atomic relaxation compared with any single-step evolution, providing a clear and minimal realization of the quantum Pontus-Mpemba effect. The analysis clarifies its physical origin in terms of the interplay between coherent Rabi oscillations, dissipative photon loss, and slow/fast decaying atom-photon states. These results not only pave the way for feasible experimental observation in both optical and circuit QED platforms, but also open new avenues for controlling and optimizing quantum relaxation processes, with potential applications in quantum technologies, fast state preparation, and nonequilibrium thermodynamics -- highlighting the broader relevance of the quantum Pontus-Mpemba effect in a rapidly evolving and highly topical area of research.

\section{Dissipative Jaynes--Cummings model}
The starting point of our analysis is provided by a dissipative version of the Jaynes--Cummings model, which provides the fundamental description of the coherent interaction between a single two-level atom and a single mode of a quantized electromagnetic field [Fig.1(a)] in the weak coupling regime \cite{JC1,JC2,JC3,JC4,DJC1}. This model can be implemented in several mature experimental platforms, such as in cavity QED and circuit QED architectures using single Rydberg atoms or trapped ions coupled to high-finesse optical cavities or  superconducting qubits coupled to coplanar-waveguide resonators. The two-level atom (qubit) is characterized by a ground state \(|g\rangle\) and an excited state \(|e\rangle\), separated in energy by \(\hbar\omega_a\). The lowering and raising operators are defined as
$\sigma^- = |g\rangle\langle e|$, $ \sigma^+ = |e\rangle\langle g|$,
and the population operator is \(\sigma_z = |e\rangle\langle e| - |g\rangle\langle g|\). The cavity mode of frequency \(\omega_c\) close to $ \omega_a$ is described by the bosonic annihilation and creation operators \(a\) and \(a^\dagger\), satisfying the usual commutation relation \([a,a^\dagger]=1\). In the dipole and rotating-wave approximation,
the coherent atom-cavity dynamics are governed by the Jaynes--Cummings Hamiltonian (we take $\hbar=1$),
\begin{equation}
H = \omega_c\, a^\dagger a + \frac{\omega_a}{2}\,\sigma_z + g\,(a^\dagger\sigma^- + a\,\sigma^+),
\label{eq:JC_Hamiltonian}
\end{equation}
where \(g \ll \omega_a \) denotes the coupling strength between the atomic transition and the cavity field. Moving to a frame rotating at the cavity frequency \(\omega_c\), the Hamiltonian becomes
\begin{equation}
H_{\mathrm{JC}} = \frac{\Delta}{2}\,\sigma_z + g\,(a^\dagger\sigma^- + a\,\sigma^+),
\label{eq:HJC_rot}
\end{equation}
where \(\Delta = \omega_a - \omega_c\) is the atom--cavity detuning. The total excitation number \(N = a^\dagger a + \sigma^+\sigma^-\) is conserved by \(H_{\mathrm{JC}}\), so that the Hilbert space decomposes into invariant subspaces of fixed excitation number.
In a realistic cavity QED setting \cite{DJC1,DJC2}, both the cavity field and the atom are coupled to dissipative environments. Photons leak out of the cavity at a rate \(\kappa\), and the atom undergoes spontaneous emission into external modes at a rate \(\gamma\). Under the Born--Markov and secular approximations, and assuming zero temperature of the reservoirs, the system's reduced density operator \(\rho(t)\) satisfies the phenomenological Lindblad master equation \cite{DJC3,Carmichael1991,DJC4,DJC4b,DJC5,DJC6,Carmichael2015,DJC7,DJC8}
\begin{equation}
\frac{ d {\rho}}{dt} = -i[H_{\mathrm{JC}}, \rho]
+ \kappa \,\mathcal{D}[a]\rho 
+ \gamma\,\mathcal{D}[\sigma^-]\rho \equiv \mathcal{L} \rho(t),
\label{eq:LME}
\end{equation}
where the dissipator is defined as $\mathcal{D}[L]\rho = L\rho L^\dagger - \frac{1}{2}\{L^\dagger L,\rho\}.$

Equation~\eqref{eq:LME} describes the interplay between coherent atom--photon exchange, atom spontaneous emission and irreversible loss of photons into the environment. In typical cavity or circuit QED experiments, the spontaneous emission of the atom into non-cavity modes is negligible compared to the photon leakage rate (\(\gamma \ll \kappa\)). Also, in the following analysis we will focus our attention to the near resonant case \( |\Delta / g| \ll 1\) of the Jaynes-Cummnig model. Finally, it should be mentioned that, for a spectrally flat photon reservoir at zero temperature, the phenomenological dissipative model described by Eq.(3) is consistent, both in the weak ($\kappa \ll g \ll \omega_a$) and strong ($g \ll \kappa \ll \omega_a$) dissipative regimes, with more accurate microscopic models \cite{DJC6}.\\ 
The most relevant observables are the atomic excitation probability,
\begin{equation}
P_e(t) = \mathrm{Tr}\!\left(\,|e\rangle\langle e|\,\rho(t)\right)
= \mathrm{Tr}\!\left(\sigma^+\sigma^-\,\rho(t)\right),
\end{equation}
and the intracavity photon number,
\begin{equation}
n_{\mathrm{ph}}(t) = \mathrm{Tr}\!\left(a^\dagger a\,\rho(t)\right).
\end{equation}
In  the $N=1$ excitation sector and in the absence of dissipation (\(\kappa=\gamma=0\)), the excitation coherently oscillates between atom and cavity mode with Rabi frequency \(\Omega_R = 2g\). The resulting \emph{vacuum Rabi oscillations} represent the hallmark of strong light--matter coupling. In the dissipative case and in the single-excitation sector, the Hilbert space is spanned by the three states 
\[
|1\rangle = |e,0\rangle, \qquad |2\rangle = |g,1\rangle, \qquad |3\rangle = |g,0\rangle,
\]
where \(|e\rangle\) and \(|g\rangle\) denote the excited and ground states of the atom, respectively, and the second label indicates the photon number in the cavity. The density matrix elements in this basis are defined as $\rho_{n,m}(t) = \langle n | \rho(t) | m \rangle$, $n,m=1,2,3$.
Starting from the Lindblad master equation (3), one obtains closed set of dynamical equations for the $ 9 $ matrix elements \(\rho_{n,m}\) in the \(|1\rangle,|2\rangle,|3\rangle\) basis, with the Liouvillian superoperator $\mathcal{L}$ being represented by a $9 \times 9$ dynamical matrix.
For the diagonal elements (populations) one has
\begin{align}
\dot{\rho}_{1,1} &= -i g (\rho_{2,1} - \rho_{1,2}) - \gamma \rho_{1,1}, \label{eq:r11}\\
\dot{\rho}_{2,2} &= +i g (\rho_{2,1} - \rho_{1,2}) - \kappa \rho_{2,2}, \label{eq:r22}\\
\dot{\rho}_{3,3} &= \gamma \rho_{1,1} + \kappa \rho_{2,2}, \label{eq:r33}
\end{align}
together with the normalization condition $
\rho_{1,1} + \rho_{2,2} + \rho_{3,3} = 1.$
The coherence between the atomic and photonic excitations evolves according to
\begin{equation}
\dot{\rho}_{1,2} = -i\Delta\,\rho_{1,2} - i g (\rho_{2,2} - \rho_{1,1}) - \frac{\gamma + \kappa}{2} \rho_{1,2},
\label{eq:r12}
\end{equation}
with \(\rho_{2,1} = \rho_{1,2}^*\). Finally, the coherences involving the vacuum state \(|3\rangle\) satisfy
\begin{align}
\dot{\rho}_{1,3} &= -i\frac{\Delta}{2} \rho_{1,3} - i g\,\rho_{2,3} - \frac{\gamma}{2}\rho_{1,3}, \label{eq:r13}\\
\dot{\rho}_{2,3} &= +i\frac{\Delta}{2} \rho_{2,3} - i g\,\rho_{1,3} - \frac{\kappa}{2}\rho_{2,3}, \label{eq:r23}
\end{align}
with \(\rho_{3,1} = \rho_{1,3}^*\) and \(\rho_{3,2} = \rho_{2,3}^*\). The relaxation dynamics is fully captured by the eigenvalues and corresponding left/right eigenvectors of the Liouvillan superoperator $\mathcal{L}$, which are given in the Appendix A.  The extension to the $N$-excitation manifold will be also briefly discussed, with the dynamical equations of the density matrix elements being given in the Appendix B. {\color{black} The effects of small thermal excitations are finally considered in the Appendix C}.
 
\begin{figure}
	\centering
	\includegraphics[width=.4\textwidth]{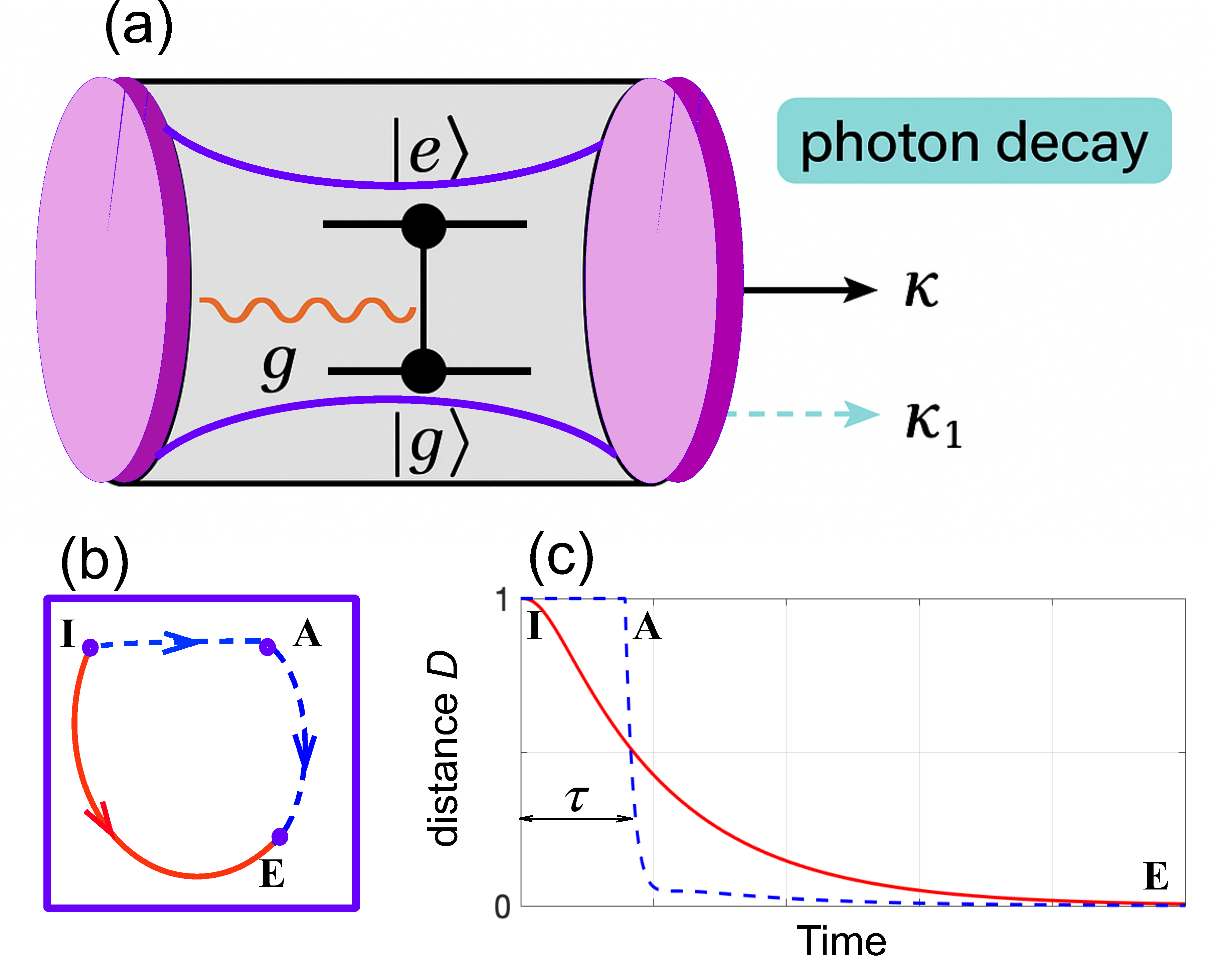}
	\caption{(a) Schematic of the dissipative Jaynes-Cummings model. A two-level atom is placed inside a high-$Q$ optical cavity with photon loss rate $\kappa$. The atom-photon coupling rate is $g$. 
	At the initial time, the atom is in the excited state while the cavity field is in the vacuum state (state I).  The dissipative dynamics drives the system toward the equilibrium state E, corresponding to the atom in the ground state and no photon in the cavity. The relaxation toward equilibrium E is quantitatively monitored by a distance measure $D(t)$ between the instantaneous density matrix $\rho(t)$ and the equilibrium state $\rho^{(E)}$.
(b,c) Illustration of the quantum Pontus-Mpemba effect. The initial atom-photon state I can relax to the equilibrium state E via two distinct pathways, which can be represented in the phase space $\mathbf{r}$ [solid red and dashed blue curves in panel (b)]. In the first scenario, the cavity loss rate is constant and large ($\kappa$, strong dissipative regime), and the system relaxes from I to E at the slow Purcell-enhanced rate, as shown by the solid red curve in panel (c).  
In the second scenario, a two-step protocol is employed: during the first stage, from $t=0$ to $t=\tau$ (half of the vacuum Rabi period), the cavity decay rate is kept small ($\kappa_1 \ll g$), allowing coherent evolution of the initial state I into an {\color{black} intermediate} state A, with the atom in the ground state and one photon in the cavity. At $t>\tau$, the cavity decay rate is switched to the same large value $\kappa$ as in the first scenario. During the interval $0<t<\tau$, the distance from equilibrium remains nearly constant, but following the cavity $Q$ switching, the relaxation toward equilibrium becomes extremely fast [dashed blue cuve in (c)], resulting in a faster approach to equilibrium compared to the first scheme .}
\end{figure}

\section{Emergence of the quantum Pontus--Mpemba effect}

In this section, we demonstrate the emergence of the quantum Pontus--Mpemba effect in a two-step dissipation protocol applied to the dissipative Jaynes-Cumming model. The protocol relies on a temporary modification of the cavity loss rate $\kappa$, i.e. quality cavity factor $Q$, to accelerate atomic relaxation compared with a single-step constant-dissipation evolution; see Figs.1(b) and (c) for a schematic.  We focus primarily on the $N=1$ excitation sector, the general case of $N$ excitations being briefly discussed at the end of the section. 
As an initial state I, we consider the atom initially prepared in its excited state while the cavity field is in the vacuum state, i.e.,
\[
\rho(0) = |1\rangle \langle 1|, \qquad \rho_{1,1}(0)=1, \quad \rho_{n,m}(0)=0 \ \text{for}\ (n,m)\neq(1,1).
\]
\begin{figure}
	\centering
	\includegraphics[width=.6\textwidth]{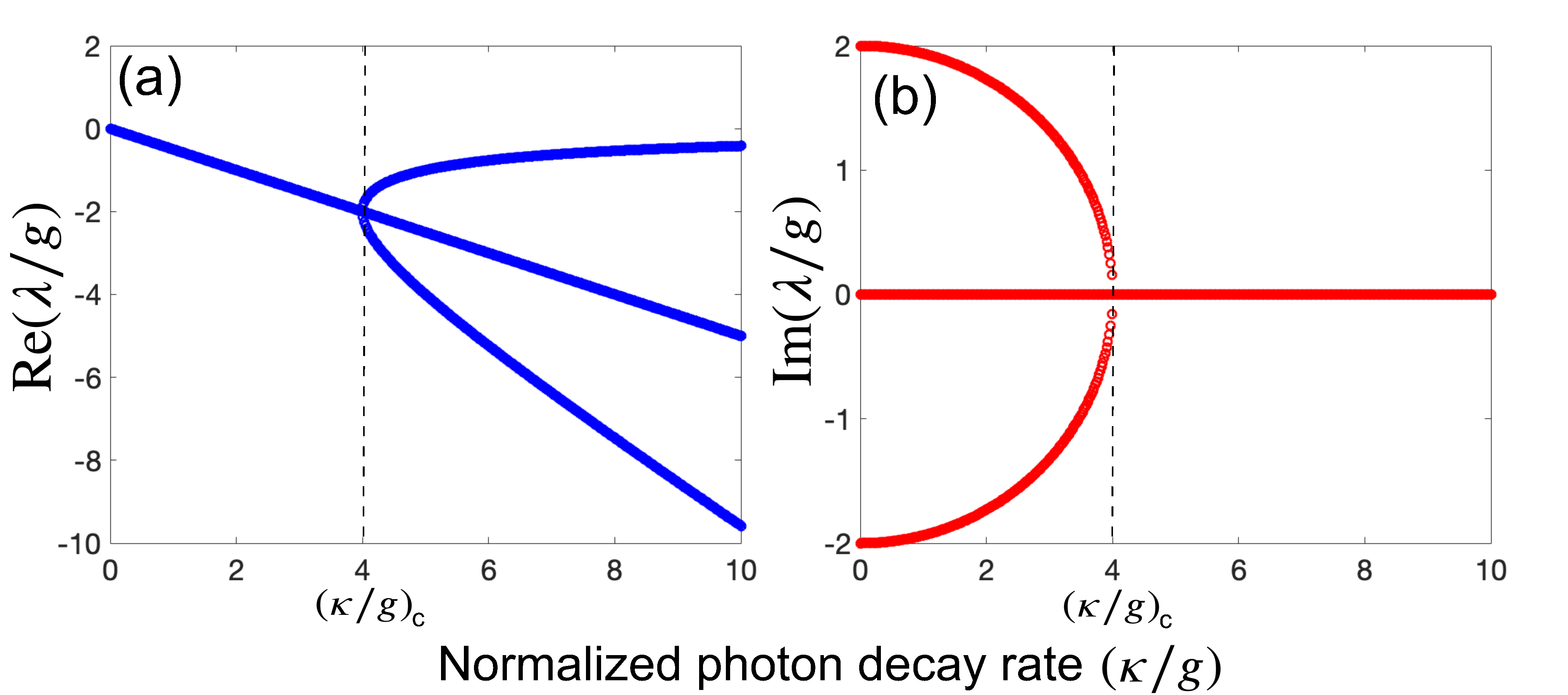}
	\caption{Behavior of (a) the real, and (b) the imaginary parts of the four eigenvalues $\lambda$ of the dynamical matrix $\mathbf{R}$ versus the normalized photon loss rate $\kappa /g$ for $\Delta=\gamma=0$. At the critical dissipation $(g/\kappa)_c=4$ two eigenvalues of $\mathbf{R}$, together with their corresponding eigenvectors, coalesce, corresponding to an exceptional point. The exceptional point separates the weak and strong dissipative regimes.}
\end{figure}
The dissipative dynamics drives the atom--cavity state toward the equilibrium state E
\begin{equation}
\rho^{(E)} = |3\rangle\langle 3|
\end{equation}
corresponding to the atom in the ground state and no photons in the cavity.
The relevant dynamics is captured by the populations $\rho_{1,1}(t)$, $\rho_{2,2}(t)$ and the coherence $\rho_{1,2}(t)=\rho_{2,1}^*(t)$, since $\rho_{3,3}(t)=1-\rho_{1,1}(t)-\rho_{2,2}(t)$ and $\rho_{1,3}(t)=\rho_{2,3}(t)\equiv 0$ for the given initial condition.
To express the dynamics compactly, we define the vector
\begin{equation}
\mathbf{r}(t) = (\rho_{1,1}(t), \rho_{2,2}(t), \rho_{1,2}(t), \rho_{2,1}(t))^\mathrm{T},
\end{equation}
so that the evolution of the relevant populations and coherence can be written as
\begin{equation}
\dot{\mathbf{r}}(t) = \mathbf{R}\, \mathbf{r}(t),
\end{equation}
where $\mathbf{R}$ is a $4\times 4$ matrix encoding the coherent atom--photon coupling $g$,  cavity decay rate $\kappa$, detuning $\Delta$ and spontaneous emission rate $\gamma$. The eigenvalues and left/right eigenvectors of $\mathbf{R}$ determine the characteristic relaxation modes of the system, and their full expressions are reported in Appendix~\ref{app:eigen} in the resonant and negligible spontaneous emission limit ($\Delta = \gamma = 0$). The behavior of the real and imaginary parts of the eigenvalues versus normalized dissipation rate $ \kappa /g$ for $\Delta=\gamma=0$ is shown in Fig.2. Note that, below the critical value $(\kappa /g)_c=4$, all eigenvalues have the same decay rate $- \kappa/2$, vanishing in the weak dissipation regime $ \kappa /g \rightarrow 0$. On the other hand, when $ \kappa /g$ is above the critical value $(\kappa /g)_c=4$, there are three distinct decay rates. At the critical point $(\kappa /g)_c=4$ there is an exceptional point and the matrix $\mathbf{R}$ becomes defective. Of particular relevance for the relaxation dynamics is the strong dissipative regime $ \kappa/g \gg 1$, which corresponds to the presence of a slow-decaying and a fast-decaying modes: in this regime the decay rate of the slowest decaying mode vanishes as $ \sim g^2 / \kappa$ whereas the decay rate of the fastest decay mode takes the asymptotic value $ \simeq - \kappa$. Interestingly, the slow and fast eigenmodes strongly overlap with the states $|1 \rangle \langle1|$ and $|2 \rangle \langle2|$, respectively. The slowest mode, with eigenvalue $\lambda \simeq -4 g^2/\kappa$, describes the Purcell-enhanced atomic decay, where an atom initially in excited state undergoes near-exponential spontaneous emission into the cavity mode.  On the other hand,  in the weak dissipation regime $\kappa \ll g$, the dynamics is dominated by coherent vacuum Rabi oscillations, which are weakly damped. 

The relaxation dynamics in phase space $\mathbf{r}$ can be schematically represented by a curve connecting the initial state I with the equilibrium final state E, as schematically shown in Fig.1(b). To quantitatively monitor relaxation toward equilibrium E, we introduce a distance measure between the instantaneous density matrix $\rho(t)$ and the equilibrium state $\rho^{(E)}$. Two common choices in opena quantum systems {\color{black}(see e.g. \cite{r12b})
}are the trace distance,
\begin{equation}
D_\mathrm{tr}(t) = \frac{1}{2} \mathrm{Tr} \, |\rho(t) - \rho^{(E)}|,
\end{equation}
and the Hilbert--Schmidt distance,
\begin{equation}
D_\mathrm{HS}(t) = \sqrt{\mathrm{Tr} \big[ (\rho(t) - \rho^{(E)})^2 \big]}.
\end{equation}
Both measures vanish when the system reaches equilibrium and provide a clear quantitative characterization of the relaxation speed. {\color{black} It should be mentioned  that quantifying the distance $D$ from equilibrium could be a nontrivial issue and a poor choice of  $D$ could
result in misleading outcomes,  as highlighted in several studies of Mpemba-like phenomena \cite{r9c,r11,r12b}.
In the present quantum open system, the quantum state is described by a full density matrix including both populations and coherences, so classical thermomajorization arguments \cite{r11} do not directly apply. On the other hand, the trace distance is a standard, contractive metric in open quantum system theory, providing a natural and consistent way to compare the relative relaxation speeds of different protocols \cite{r12b}. The Hilbert-Schmidt distance is sometimes used for convenience, but it is not generally contractive under quantum dynamics. 
While other operational or thermodynamic-oriented metrics, such as the quantum relative entropy, can be formally defined, they are generally not suited for quantifying relaxation, as they may diverge if the support of $\rho(t)$ is not contained in that of $\rho^{(E)}$. }

\begin{figure}[t]
\centering
\includegraphics[width=0.8\textwidth]{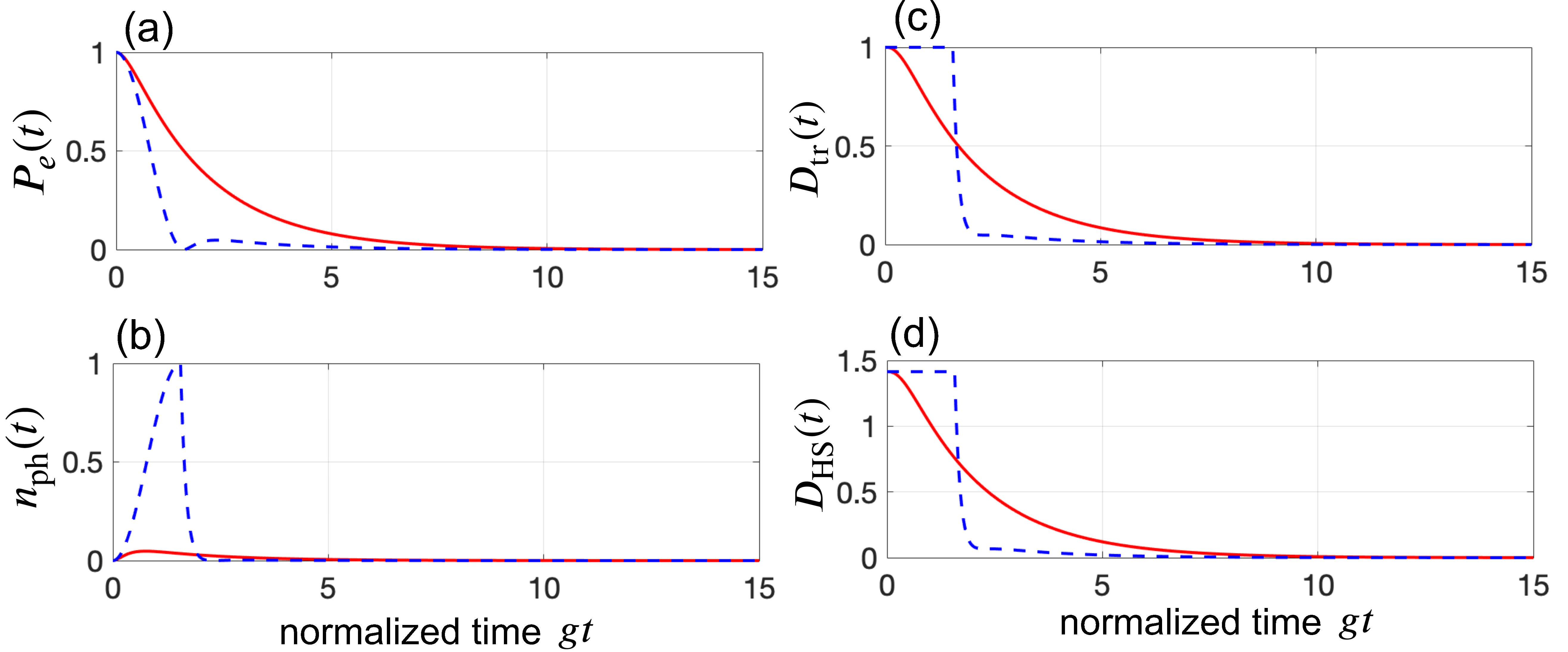}
\caption{
Typical dynamical evolution in the dissipative Jaynes--Cummings system illustrating the quantum Pontus--Mpemba effect. Dashed blue lines correspond to the two-step protocol, where the cavity decay rate is initially low ($\kappa_1 \ll g$) and switched to high loss ($\kappa \gg g$) at time $\tau = \pi/(2 g)$. Solid red lines correspond to the single-step high-loss evolution. 
(a) Excited atomic population $P_e(t)=\rho_{1,1}(t)$.  
(b) Cavity photon number $n_{ph}(t)=\rho_{2,2}(t)$.  
(c,d) Distance from equilibrium (either trace distance $D_\mathrm{tr} = \frac{1}{2}\mathrm{Tr}|\rho(t)-\rho^{(E)}|$ or Hilbert--Schmidt distance $D_\mathrm{HS} = \sqrt{\mathrm{Tr}[(\rho(t)-\rho^{(E)})^2]}$).  
The two-step protocol leads to accelerated relaxation: the atomic population decays faster and the system reaches equilibrium earlier than in the single-step high-loss evolution. Parameter values used in the numerical simulations are $\kappa/g=8$, $\kappa_1 /g=0$, and $\Delta=\gamma=0$.}
\end{figure}

\begin{figure}[t]
\centering
\includegraphics[width=0.8\textwidth]{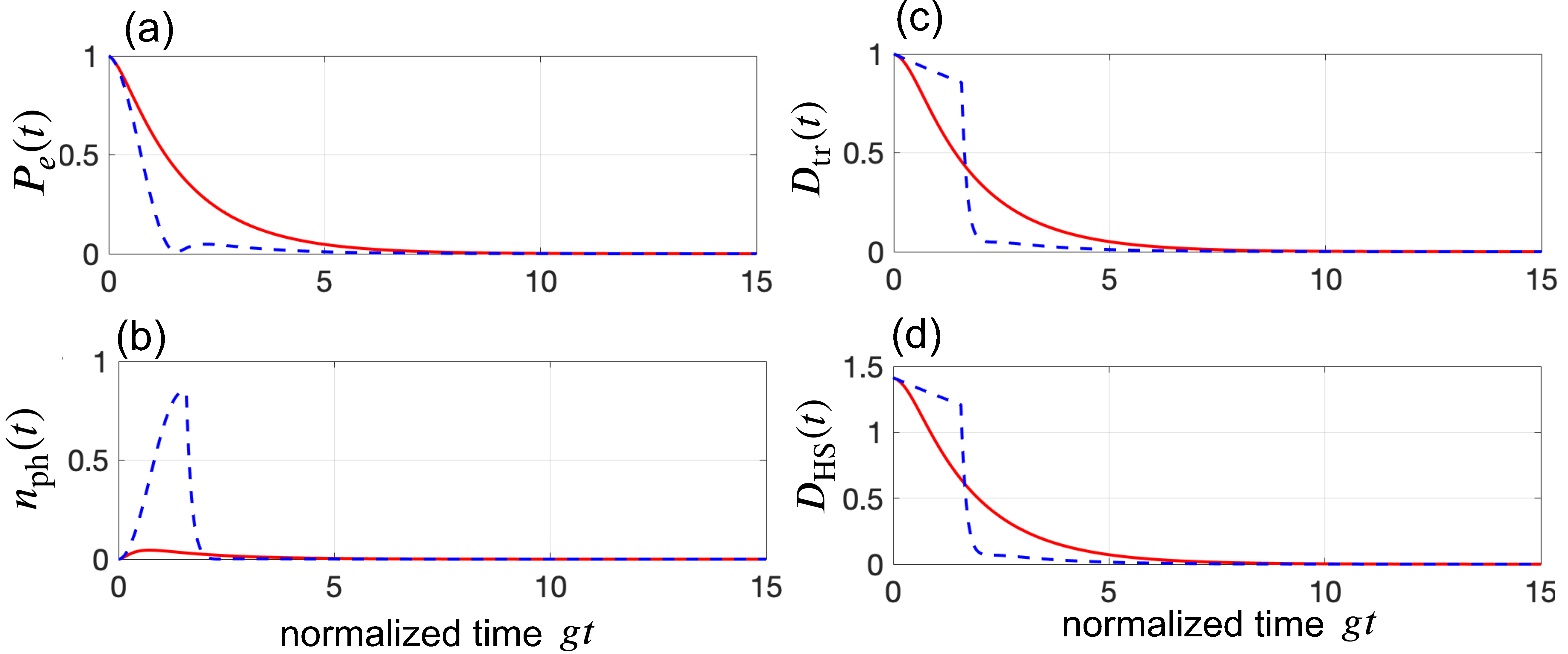}
\caption{
Same as Fig.3, but for parameter values $\kappa/g=8$, $\kappa_1 /g=0.1$, $\Delta/g=-0.2$, and $\gamma / g=0.1$.}
\end{figure}

\begin{figure}[t]
\centering
\includegraphics[width=0.95\textwidth]{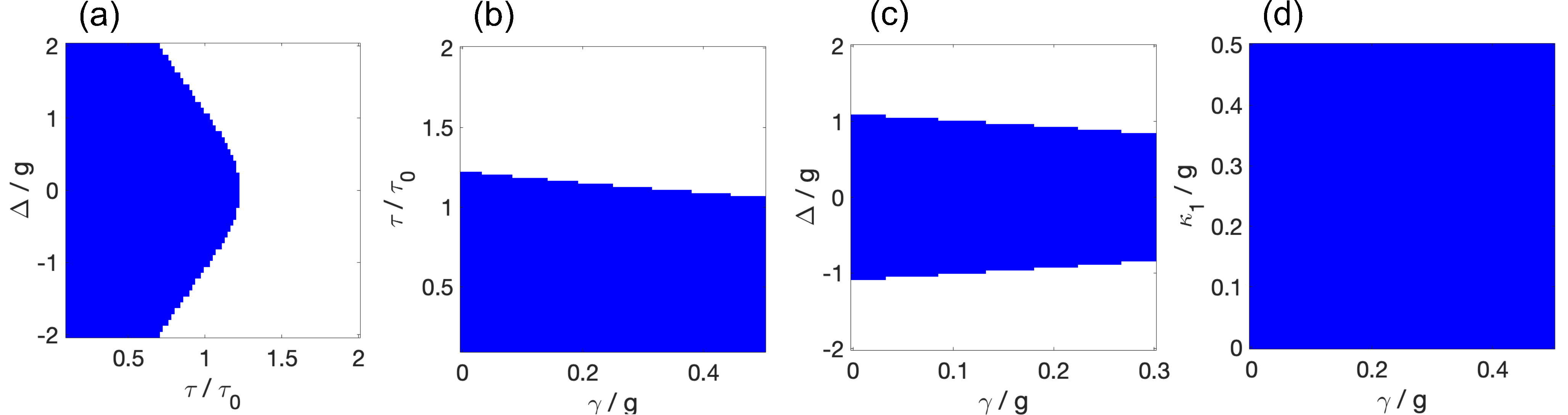}
\caption{{\color{black}Phase diagrams showing the regions of parameter space in which the Pontus--Mpemba effect is observable. The dark shaded areas indicate where the effect occurs, while the white regions correspond to its absence. (a) Domain in the $(\tau/\tau_0,\, \Delta/g)$ plane for $\kappa/g = 8$, $\kappa_1/g = 0$, and $\gamma = 0$. (b) Domain in the $(\tau/\tau_0,\, \gamma/g)$ plane for $\kappa/g = 8$, $\kappa_1/g = 0$, and $\Delta = 0$. (c) Domain in the $(\gamma/g,\, \Delta/g)$ plane for $\kappa/g = 8$, $\kappa_1/g = 0$, and $\tau/\tau_0 = 1$. (d) Domain in the $(\gamma/g,\, \kappa_1/g)$ plane for $\kappa/g = 8$, $\Delta = 0$, and $\tau/\tau_0 = 1$.}}
\end{figure}

\begin{figure}[t]
\centering
\includegraphics[width=0.8\textwidth]{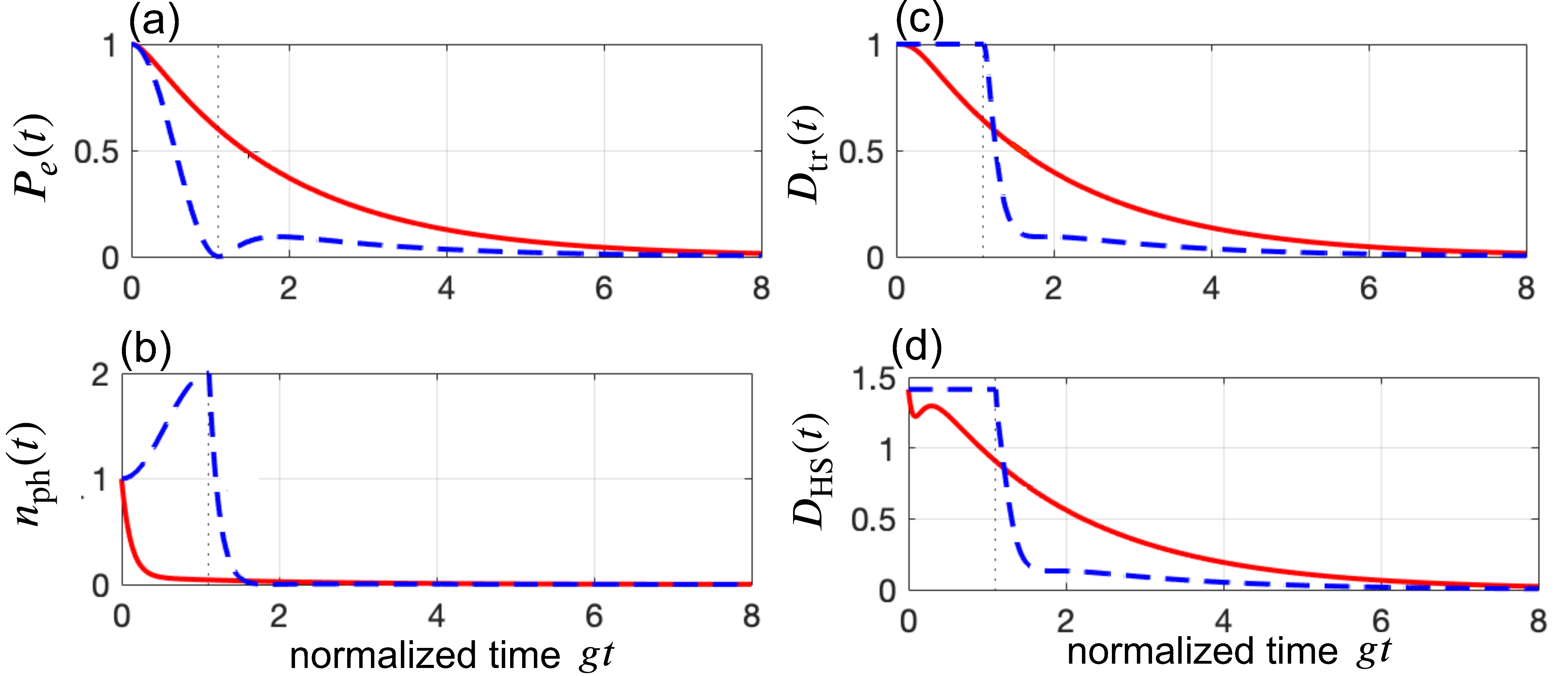}
\caption{
Same as Fig.3, but for the initial state I given by $|e,1 \rangle \langle e,1|$. The swtiching time is $\tau=\pi/(2 \sqrt{}2g)$.}
\end{figure}

The quantum Pontus--Mpemba effect manifests itself when comparing the relaxation dynamics under single-step versus two-step cavity dissipation protocols, as schematically shown in Figs.1(b) and (c).  In the single-step protocol,  the cavity quality factor is kept low (strong dissipation regime), corresponding to a high photon loss rate $\kappa  \gg g$. In this regime, the system evolves directly toward the equilibrium state $\rho^{(E)} = |3\rangle\langle 3|$ at the rate set by the slow-decaying eigenmode of the dynamical matrix $\mathbf{R}$; see solid red curves in Figs.1(b) and (c).  Physically, this occurs because the initial state $\rho(0) = |1\rangle \langle 1|$ does not populate the fast-decaying cavity mode: the system starts with zero photons, so the fast cavity mode remains unexcited. As a result, the atomic decay is essentially governed by the slow Purcell-enhanced channel with rate $\sim 4 g^2 / \kappa$ (see Appendix A for more technical details).

In the second scenario [dashed blue curves in Figs.1(b) and (c)], the cavity is initially set to a high quality factor, corresponding to a small loss rate $ \kappa_1 \ll g$, and is maintained in this configuration for a short time $\tau$ before switching to the larger loss rate $\kappa$. The duration $\tau$ is chosen to correspond to approximately half of the vacuum Rabi period, i.e., 
$\tau \simeq \tau_0 \equiv \frac{\pi}{2 g}$,
so that the initial state $\rho(0) = |1\rangle \langle 1|$, with atomic excitation and no photon in the cavity, is coherently driven close to the {\color{black} intermediate state A of the Hilbert space}, $\rho(t= \tau) \simeq |2\rangle \langle 2|$, corresponding to one photon in the cavity and the atom in the ground state. During this short coherent evolution, the slow dissipation ensures that the atomic excitation is almost entirely transferred into the photon field without significant loss, and thus the distance $D(t)$ remains constant. At time $t= \tau$, after switching the cavity back to the high-loss regime $\kappa$, the photon in the cavity rapidly decays [dashed blue curve connecting states A and E in Figs.1(c)], effectively accelerating the overall atomic relaxation compared with the single-step high-loss scenario.  This is because the {\color{black} intermediate} state A strongly overlaps with the fast decaying mode with decay rate $\lambda_4 \sim  -\kappa$. In other words, by temporarily allowing coherent Rabi oscillations in a low-loss cavity, the system selectively excites the fast-decaying mode, which then rapidly dumps energy when the cavity is switched back to the high-loss regime. This two-step evolution demonstrates a clear and very simple manifestation of the quantum Pontus--Mpemba effect: the atom relaxes faster through the quenched protocol than it would under constant high dissipation. {\color{black}  At a conceptual level, the accelerated relaxation arises because the initial coherent stage steers the system into an intermediate state with negligible overlap with the slowest Liouvillian mode, thus opening a fast relaxation channel when the cavity loss is increased; this mechanism is closely related to the mode-selection principle underlying strong Mpemba effects and shortcut-to-equilibration strategies \cite{r13,r14}, with the important distinction that in the Pontus--Mpemba protocol one must also account for the finite time~\(\tau\) required to drive the initial state~\(I\) into the intermediate state~\(A\) \cite{P1}.}

{\color{black} It is worth commenting on the role of complex Liouvillian eigenvalues and the associated oscillatory relaxation dynamics, which could give rise to multiple Mpemba effect \cite{Referee1,Referee2}. As discussed in Appendix~A, the dissipative Jaynes--Cummings Liouvillian exhibits an exceptional point at a critical value of the cavity decay rate, where two eigenvalues coalesce and the system transitions from an underdamped regime with complex--conjugate eigenvalues to an overdamped regime with two distinct real decay rates. The Pontus--Mpemba effect reported here occurs in the overdamped region above the EP, where the separation between the slow and fast decay rates becomes pronounced. The two-step protocol exploits precisely this spectral separation, and therefore the presence of oscillatory decay below the EP does not qualitatively influence the occurrence of the effect in the parameter regime studied.}

 Figure~3 illustrates the key features of the quantum Pontus--Mpemba effect. The two-step protocol (dashed blue lines) accelerates the decay of the atomic population and drives the system to equilibrium faster than the single-step high-loss evolution (solid red lines). The cavity photon population initially rises during the low-loss stage, allowing the fast-decaying mode to be populated and subsequently dissipated when the cavity is switched to high loss. The distance from equilibrium confirms that the two-step evolution achieves faster thermalization. {\color{black} It should be mentioned that, while the photon number and the excited-state population do not fully characterize the quantum state, they can be regarded as practical ``good observables'' for detecting the Mpemba effect~\cite{arxiv:2512.02709}. Indeed, as shown in Figs.~3(a,b), the Pontus-Mpemba effect is clearly visible in the decay dynamics of these quantities. These observables are experimentally accessible and can be measured without performing full quantum state tomography, providing a reliable and convenient signature of the effect in realistic cavity QED setups.}

{\color{black}
The above analysis assumes exact zero atom--photon detuning $\Delta = 0$, negligible spontaneous emission of the atom $\gamma = 0$, and a switching time $\tau = \tau_0$ equal to half of the vacuum Rabi period. However, it should be emphasized that the Pontus--Mpemba effect is not a resonance phenomenon; rather, it is remarkably robust and can be observed over a wide range of parameters, as illustrated in Figs.~4 and 5.
Figure~4 shows an example of the relaxation dynamics when including a small spontaneous emission rate $\gamma$ and a small atom--photon detuning $\Delta$. As one can see, the two-step protocol continues to accelerate relaxation. The robustness of the quantum Pontus--Mpemba effect in the dissipative Jaynes--Cummings model is further demonstrated in Fig.~5, where the shaded regions in parameter-space plots highlight the parameter values for which the Pontus--Mpemba effect is observed, while the white regions correspond to the absence of the effect.
The diagrams are obtained by comparing the relaxation dynamics in the two protocols and evaluating a fixed final observation time $t^* > \tau$ (typically we set $t^* = 8/g$). The Pontus--Mpemba effect is deemed to occur when the trace distance $D_{\mathrm{tr}}(t^*)$ at time $t^*$ is smaller in the two-step protocol than in the single-step protocol. Similar diagrams are obtained by considering the Hilbert--Schmidt distance. These results clearly indicate that the Pontus--Mpemba effect is a robust phenomenon, potentially observable under realistic experimental conditions subject to unavoidable imperfections.
}

The previous analysis focused on the case of a single initial excitation ($N=1$), for which simple analytical expressions for the relaxation dynamics can be obtained. However, the effect is expected to persist when the system is initially prepared with $N \geq 2$ excitations. For instance, consider the pure initial state $\rho(0)=|e, N-1 \rangle \langle e, N-1|$ (state I), where the atom is excited and the cavity contains $(N-1)$ photons. 
Under dissipation, the system relaxes toward the ground equilibrium state $\rho^{(E)}=|g,0 \rangle \langle g,0|$ (state E). Nevertheless, switching the cavity decay rate can strongly influence this relaxation process. In the strongly dissipative regime with a constant decay rate $\kappa \gg g$, relaxation is slowed down because the atomic transition toward the ground state becomes inefficient. In contrast, if the cavity decay rate is initially kept small (or ideally negligible), $\kappa_1 \ll g$, for a time interval $\tau$ equal to half of the Rabi oscillation period in the $N$-excitation manifold, i.e. $\tau = \pi / (2 g \sqrt{N})$, the system evolves coherently from the initial state I toward the {\color{black}intermediate} pure state A, $|g, N \rangle \langle g, N|$. Once the cavity loss rate is switched to a large value, this {\color{black}intermediate} state rapidly relaxes to the equilibrium state E. 
{\color{black}We note that the relative speed-up provided by the two-step protocol does not significantly increase with $N$. For large cavity decay rates $\kappa$, the photon population in the first stage rapidly relaxes to zero on a timescale $\sim 1/\kappa$, effectively projecting the dynamics onto the single-excitation manifold before the atomic relaxation takes place. Consequently, the subsequent evolution closely mirrors the $N=1$ case, and the benefit of the two-step protocol saturates already for small excitation numbers. The very large-$N$ regime, where $\sqrt{N}\, g$ can exceed $\kappa$ and lead to damped Rabi oscillations or non-monotonic relaxation, lies beyond the scope of this work but may display richer Mpemba-like behavior.}
Explicit dynamical equations of the density matrix elements in the $N$ excitation manifold are given in Appendix~B. Figure~5 shows a representative example of the relaxation dynamics, obtained by numerically solving the Lindblad master equation for the initial state $\rho(0)=|e,1 \rangle \langle e,1|$ ($N=2$ excitation manifold) under both the one- and two-step protocols. In this case, the switching time $\tau$ is chosen to be half of the Rabi oscillation period in the $N=2$ excitation sector, $\tau \simeq \pi / (2 \sqrt{2} g)$. During the first stage, with low cavity losses, the initial state $|e,1 \rangle \langle e,1|$ is coherently driven into the {\color{black}intermediate} state $|g,2 \rangle \langle g,2|$, which subsequently undergoes rapid relaxation toward the equilibrium state $|g,0 \rangle \langle g,0|$ after the cavity-$Q$ switching. The numerical results clearly demonstrate the persistence of the Pontus--Mpemba effect, showing accelerated relaxation once the system is coherently driven to a state where the atom is in the ground state and two photons occupy the resonator.

Let us finally briefly discuss possible experimental platforms of the dissipative Jaynes--Cummings model that could be feasible for an experimental demonstration of the Pontus--Mpemba effect predicted in this work. The Jaynes--Cummings model with controllable parameters and operating conditions has been realized in several mature experimental platforms, ranging from cavity QED and circuit QED architectures to hybrid implementations using semiconductor quantum dots or color centers in nanophotonic cavities \cite{r61,r62,r63,r64,r65,r66,r66b,r67,r68}. In cavity QED, strong atom--photon coupling has been achieved with single Rydberg atoms or trapped ions coupled to high-finesse optical cavities \cite{JC4,r61}. Circuit QED architectures based on superconducting qubits coupled to coplanar-waveguide resonators provide an even more versatile platform, where both the atom--cavity coupling strength $g$ and the cavity decay rate $\kappa$ can be tuned in real time through flux or parametric control \cite{r62,r63,r64,r66b}. Recent advances in tunable-dissipation engineering, such as variable couplers and quantum switches, enable fast quenches of cavity loss rates \cite{r65,r66} comparable to those required for the two-step relaxation protocol underlying the Pontus--Mpemba effect discussed in this work.

In a representative superconducting resonator setup with variable coupling to a transmission line, as realized in Ref.~\cite{r64}, typical parameters are $\omega_a = 2\pi \times 6.5~{\rm GHz}$ for the qubit transition frequency and $g = 2\pi \times 12~{\rm MHz}$ for the qubit--photon coupling rate. The cavity loss rate $\kappa$ can be rapidly switched from below $g$ (low-dissipation regime, with a minimum ratio $\kappa/g \sim 0.01$ set by the intrinsic resonator decay rate) to well above $g$ (high-dissipation regime, reaching $\kappa/g \sim 3$ in Ref.~\cite{r64}) on nanosecond timescales ($\sim 2~{\rm ns}$). A larger effective ratio $\kappa/g$ could be achieved by engineering resonators with intentionally lower quality factors, for instance by using broadband Purcell filters or dynamically adjustable output couplers that enhance external coupling while keeping internal losses negligible. Alternatively, the coupling $g$ can be reduced by employing weakly coupled transmon or fluxonium qubits designed with smaller dipole moments. These strategies would allow exploration of the regime $\kappa/g > 5$ or even approaching $\kappa/g \sim 10$, where the transition from underdamped to overdamped relaxation becomes most pronounced and the Pontus--Mpemba effect is expected to be most visible.
{\color{black}The theoretical analysis assumes an idealized instantaneous quench of the cavity $Q$ factor, however this is not a limiting constraint in practical settings. In present circuit-QED setups, quench times of a few nanoseconds remain much shorter than the relevant Rabi period $\pi/g$, so that finite quench speeds introduce only minor corrections. The protocol is also robust against moderate timing jitter in the switching time $\tau$, as illustrated in Fig.~5: deviations of several percent in $\tau$ from half of the Rabi period $\tau_0 = \pi/(2g)$ still yield nearly optimal relaxation. Residual cavity losses during the low-loss stage can likewise be tolerated [Fig.~5(d)].}
{\color{black}Atomic spontaneous emission can also be safely neglected in these platforms. In circuit QED, radiative decay of the qubit into non-cavity channels is strongly Purcell-suppressed, typically giving $\gamma \sim (10~{\rm kHz} - 100~{\rm kHz}) \ll g, \kappa_{2}$, which is negligible on the timescales relevant for the protocol. Similar conditions are realized in cavity-QED systems employing long-lived Rydberg or trapped-ion states.}
{\color{black}Finally, although the cavity mode may host a small thermal photon population at dilution-refrigerator temperatures, the associated occupation number remains very small ($n_{\mathrm{th}} \lesssim 10^{-3}$ for $\omega/2\pi \sim 5$--$7~{\rm GHz}$ at $T = 15$--$20~{\rm mK}$). As shown in Appendix~C, the Pontus--Mpemba effect is robust against such small thermal backgrounds, which introduce only minor quantitative corrections to the relaxation trajectories.}

\section{Conclusion and discussion}

We have proposed and analyzed a minimal realization of the quantum Pontus--Mpemba effect in cavity quantum electrodynamics using the dissipative Jaynes--Cummings model. By restricting the dynamics to the single-excitation sector, we have shown that a sudden quench of the cavity decay rate can induce a two-step relaxation trajectory that reaches equilibrium faster than any single-step evolution under constant dissipation. The accelerated relaxation originates from the nontrivial interplay between coherent atom--photon exchange and dissipative photon loss, which reshapes the relaxation spectrum of the Liouvillian superoperator. This establishes a clear and experimentally accessible setting to observe the quantum Pontus--Mpemba effect with state-of-the-art cavity or circuit QED platforms.

Beyond serving as a proof-of-principle demonstration, these results highlight the potential of controlled dissipation and dynamical quenches as powerful tools for engineering non-equilibrium quantum relaxation. The Jaynes--Cummings platform bridges fundamental open-system physics and emerging applications in quantum technologies, including fast qubit reset, optimal state transfer, and thermodynamic control at the quantum level. The demonstration of a genuine two-step relaxation speed-up underscores the rich structure of dissipative quantum dynamics, even in zero-dimensional systems.
{\color{black}Importantly, while our analysis focuses on a two-level atom, the underlying mechanism of spectrally separated Liouvillian modes and controllable population transfer is expected to persist in more complex atom--field systems, including multilevel atoms interacting with a single cavity mode via extended Jaynes--Cummings-type Hamiltonians. This suggests that Pontus--Mpemba-type acceleration could be observed beyond the simplest two-level setting.}

Future work could extend this framework along several directions. First, an analysis in the dispersive Jaynes--Cummings regime would clarify how photon-number-dependent Stark shifts and effective Kerr nonlinearities modify the Pontus--Mpemba dynamics. Second, exploring multi-photon and collective effects could reveal cooperative or many-body realizations of the quantum Pontus--Mpemba effect, directly connecting to recent studies on the breakdown of photon blockade and dissipative quantum phase transitions \cite{Carmichael2015}. Additional extensions may include time-periodic (Floquet) modulation of the dissipation rate, coupling to non-Markovian reservoirs, or implementation in multi-cavity and hybrid architectures, where interference between relaxation pathways could further enhance or suppress the effect. From an experimental perspective, circuit-QED architectures offer a particularly promising route toward realizing the proposed scenario.

\appendix
\section{Appendix A: Eigenvalues and eigenvectors of the dynamical matrix in the $N=1$ excitation manifold}
\label{app:eigen}

In the resonant case (\(\Delta = 0\)) and neglecting atomic spontaneous emission (\(\gamma = 0\)), the dynamics of the relevant atom--cavity populations and coherence can be compactly expressed in vector form as
\[
\dot{\mathbf{r}}(t) = \mathbf{R}\, \mathbf{r}(t), 
\qquad
\mathbf{r}(t) = (\rho_{1,1}, \rho_{2,2}, \rho_{1,2}, \rho_{2,1})^\mathrm{T},
\]
where the dynamical matrix \(\mathbf{R}\) is
\[
\mathbf{R} =
\begin{pmatrix}
0 & 0 & i g & -i g \\
0 & -\kappa & -i g & i g \\
i g & -i g & -\kappa/2 & 0 \\
-i g & i g & 0 & -\kappa/2
\end{pmatrix}.
\]
This form correctly reproduces the physical limits: it is anti-Hermitian for \(\kappa = 0\), reflecting purely coherent Rabi oscillations, and Hermitian diagonal when \(g = 0\), corresponding to purely dissipative relaxation. 
The characteristic polynomial of \(\mathbf{R}\) is
\[
\det(\mathbf{R}-\lambda I) = (\lambda + \kappa/2)^2 \left( \lambda^2 + \kappa \lambda + 4 g^2 \right) = 0.
\]
The eigenvalues are therefore
\[
\lambda_{1,2} = -\frac{\kappa}{2}, \qquad
\lambda_{3,4} = -\frac{\kappa}{2} \pm \sqrt{\frac{\kappa^2}{4} - 4 g^2}.
\]
Figure~2 in the main text illustrates the real and imaginary parts of these eigenvalues as functions of the normalized dissipation rate \(\kappa/g\). 
Since \(\mathbf{R}\) is non-Hermitian, the \emph{right} and \emph{left} eigenvectors are distinct. The right eigenvectors \(\mathbf{v}_i\) satisfy
\[
\mathbf{R} \, \mathbf{v}_i = \lambda_i \mathbf{v}_i,
\]
while the left eigenvectors \(\mathbf{u}_i\) satisfy
\[
 \mathbf{R}^{\dag} \mathbf{u}_i = \lambda_i^* \mathbf{u}_i.
\]
They obey the biorthogonality condition \( \langle \mathbf{u}_i | \mathbf{v}_j \rangle= \delta_{ij}\).  Since \(\mathbf{R}\) coincides with its transpose, it follows that \(\mathbf{u}_i=\mathbf{v}_i^*\).  In the absence of exceptional points, any initial state \(\mathbf{r}(0)\) can be decomposed in the right eigenmode basis, and thus the general solution of the linear system reads
\[
\mathbf{r}(t) = \sum_{i=1}^4 \mathbf{v}_i \, \langle  \mathbf{u}_i|  \mathbf{r}(0) \rangle \, e^{\lambda_i t}.
\]
This decomposition is particularly useful for the {quantum Pontus--Mpemba effect}, as it makes explicit how the initial state can selectively populate slow or fast modes, leading to non-monotonic relaxation trajectories. The right eigenvectors of the dynamical matrix $\mathbf{R}$ corresponding to the four eigenvalues can be expressed as follows.
The degenerate eigenspace at $\lambda_{1}=\lambda_2=-\kappa/2$ is two-dimensional; a convenient
unnormalized basis is
\[
\mathbf v_{1} = \begin{pmatrix} 0 \\[2pt] 0 \\[2pt] 1 \\[2pt] 1 \end{pmatrix},  	\;\;\;
\mathbf v_{2} = \begin{pmatrix} 1 \\[2pt] 1 \\[2pt] -\dfrac{i\kappa}{2g} \\[6pt] \dfrac{i\kappa}{2g} \end{pmatrix}.
\]
For the nondegenerate eigenvalues $\lambda_{3,4}$, one obtains the explicit form for the right eigenvectors
\[
\mathbf v_{3,4} \propto
\begin{pmatrix}
1 \\[4pt]
-\dfrac{\lambda_{3,4}}{\kappa+\lambda_{3,4}} \\[6pt]
\dfrac{i g\,(\kappa+2\lambda_{3,4})}{(\kappa+\lambda_{3,4})(\lambda_{3,4}+\kappa/2)} \\[8pt]
-\dfrac{i g\,(\kappa+2\lambda_{3,4})}{(\kappa+\lambda_{3,4})(\lambda_{3,4}+\kappa/2)}
\end{pmatrix}.
\]
These vectors are unnormalized. At the critical dissipation rate $( \kappa /g)_c=4$ the two eigenvalues $\lambda_{3,4}$ and their corresponding eigenvectors $\mathbf{v}_{3,4}$ coalesce, corresponding to the emergence of an exceptional point.


To gain further physical insight, it is instructive to consider the limiting regimes of weak and strong dissipation. In these limits, the dynamical matrix \(\mathbf{R}\) becomes effectively anti-Hermitian or Hermitian, so that the left and right eigenvectors coincide. In the weak dissipation regime (\(\kappa \ll g\)), the coherent atom--photon coupling dominates. The eigenvalues \(\lambda_{3,4}\) form a complex-conjugate pair with small negative real parts, describing damped vacuum Rabi oscillations, while the degenerate eigenvalues \(\lambda_1 = \lambda_2 = -\kappa/2\) govern the slow relaxation toward equilibrium.

In the strong dissipation regime (\(\kappa \gg g\)), the cavity relaxes rapidly, effectively providing a dissipative environment for the atom. The eigenvalues \(\lambda_3\) and \(\lambda_4\) correspond to slow and fast decaying modes, with \(\lambda_3 \simeq -4 g^2 / \kappa \rightarrow 0\) and \(\lambda_4 \simeq -\kappa\) as \(\kappa / g \to \infty\). The slowest mode \(\lambda_3\) represents the Purcell-enhanced atomic decay, whereas \(\lambda_4\) describes the rapid cavity relaxation. The initial state \(I = |1\rangle\langle 1|\), has strong overlap with the slow mode \(\mathbf{v}_3 \simeq (1,0,0,0)^\mathrm{T}\), leading to slow relaxation in a single-step protocol. Conversely, the {\color{black} intermediate} state $A=|2 \rangle \langle 2|$ strongly overlaps with the fast mode \(\mathbf{v}_4 \simeq (0,1,0,0)^\mathrm{T}\), producing rapid relaxation after cavity \(Q\)-switching.

Overall, this eigenmode analysis clarifies the interplay between coherent atom--photon exchange and cavity dissipation and highlights the hierarchy of timescales that underlies the accelerated relaxation observed in the two-step cavity decay protocol discussed in the main text.

\section*{Appendix B: Density-matrix equations in the $N$-excitation manifold}

We consider the dissipative Jaynes--Cummings model restricted to the manifold with at most $N$ total excitations. 
The truncated Hilbert space is spanned by the states
\[
\mathcal{B}_N = 
\bigl\{
    |g,0\rangle, |g,1\rangle, \dots, |g,N\rangle,
    |e,0\rangle, |e,1\rangle, \dots, |e,N-1\rangle
\bigr\},
\]
so that the total dimension is $2N+1$.
We denote the density-matrix elements in this basis as
\[
\rho^{g,g}_{n,m} = \langle g,n | \rho | g,m \rangle, \quad
\rho^{e,e}_{n,m} = \langle e,n | \rho | e,m \rangle, \quad
\rho^{e,g}_{n,m} = \langle e,n | \rho | g,m \rangle, \quad
\rho^{g,e}_{n,m} = \langle g,n | \rho | e,m \rangle.
\]
with indices in the ranges
\[
\begin{aligned}
&\rho^{g,g}_{n,m}: n,m=0,\dots,N,\\
&\rho^{e,e}_{n,m}: n,m=0,\dots,N-1,\\
&\rho^{e,g}_{n,m}: n=0,\dots,N-1,\; m=0,\dots,N,\\
&\rho^{g,e}_{n,m}: n=0,\dots,N,\; m=0,\dots,N-1.
\end{aligned}
\]
The Hamiltonian in the truncated manifold is
\[
H_{JC} = -\frac{\Delta}{2}\sum_{n=0}^{N}|g,n\rangle\langle g,n|
         +\frac{\Delta}{2}\sum_{n=0}^{N-1}|e,n\rangle\langle e,n|
         + g\sum_{n=1}^{N}\sqrt{n}\bigl(|g,n\rangle\langle e,n-1| + |e,n-1\rangle\langle g,n|\bigr),
\]
with detuning $\Delta$ and coupling $g$. From the Lindblad master equation (3), it follows that 
the density-matrix elements evolve according to the set of equations
\[
\begin{aligned}
\dot{\rho}^{g,g}_{n,m} &=
  - i g \left(\sqrt{n}\,\rho^{e,g}_{\,n-1,m} - \sqrt{m}\,\rho^{g,e}_{\,n,m-1}\right)
  + \kappa \left(\sqrt{(n+1)(m+1)}\,\rho^{g,g}_{\,n+1,m+1} - \tfrac{n+m}{2}\,\rho^{g,g}_{n,m}\right)
  + \gamma\,\rho^{e,e}_{n,m},\\[1ex]
\dot{\rho}^{e,e}_{n,m} &=
  - i g \left(\sqrt{n+1}\,\rho^{g,e}_{\,n+1,m} - \sqrt{m+1}\,\rho^{e,g}_{\,n,m+1}\right)
  + \kappa \left(\sqrt{(n+1)(m+1)}\,\rho^{e,e}_{\,n+1,m+1} - \tfrac{n+m}{2}\,\rho^{e,e}_{n,m}\right)
  - \gamma\,\rho^{e,e}_{n,m},\\[1ex]
\dot{\rho}^{e,g}_{n,m} &=
  i \Delta\,\rho^{e,g}_{n,m} 
  - i g \left(\sqrt{n+1}\,\rho^{g,g}_{\,n+1,m} - \sqrt{m}\,\rho^{e,e}_{\,n,m-1}\right)
  + \kappa \left(\sqrt{(n+1)(m+1)}\,\rho^{e,g}_{\,n+1,m+1} - \tfrac{n+m}{2}\,\rho^{e,g}_{n,m}\right)
  - \tfrac{\gamma}{2}\,\rho^{e,g}_{n,m},\\[1ex]
\dot{\rho}^{g,e}_{n,m} &=
  - i \Delta\,\rho^{g,e}_{n,m} 
  - i g \left(\sqrt{n}\,\rho^{e,e}_{\,n-1,m} - \sqrt{m+1}\,\rho^{g,g}_{\,n,m+1}\right)
  + \kappa \left(\sqrt{(n+1)(m+1)}\,\rho^{g,e}_{\,n+1,m+1} - \tfrac{n+m}{2}\,\rho^{g,e}_{n,m}\right)
  - \tfrac{\gamma}{2}\,\rho^{g,e}_{n,m}.
\end{aligned}
\]
All terms with indices outside the allowed ranges are set to zero. 
In the presence of cavity dissipation, all trajectories relax toward the equilibrium state $\rho^{(E)} = |g,0\rangle\langle g,0|$.

{\color{black}

\section*{Appendix C: Thermal excitation effects}

In this Appendix we briefly discuss the effects of small thermal excitations of atom and photons, in contacts with thermal baths at finite temperatures, on the  relaxation dynamics and the onset of the Pontus-Mpemba effect. We consider the dissipative Jaynes--Cummings model, including small mean thermal photon number $n_\mathrm{th} \ll 1$ in the cavity and small thermal atomic excitation $n_\mathrm{th,atom} \ll 1$ arising from non-zero temperatures of the baths. While thermal atomic excitation can be readily included in the low-manifold model presented in Sec.2,  the presence of thermal photons requires care since the dynamics formally involve an infinite Hilbert space with the basis $\{ |g,n \rangle , |e,n \rangle\}$ ($n=0,1,2,3,...$) . However, in $n_\mathrm{th} \ll 1$ limit as shown in Sec.~III of Ref.~\cite{DJC2}
 restricting the Hilbert space to the three manifold $|1 \rangle \equiv |e,0\rangle$, $ |2 \rangle \equiv |g,1\rangle$, $|3 \rangle \equiv  |g,0\rangle$ can still provide an adequate approximation to capture the effects of photon thermal number excitation, without introducing a larger Hilbert space basis.

Let us start with the Lindblad master equation [Eq.(3) in the main text], extended to include thermal photons and atomic excitation due to interaction with the thermal baths at finite temperatures. The equation reads
\[
\frac{d\rho}{dt} = -i [H_{\mathrm{JC}}, \rho] 
+ \kappa (1+n_\mathrm{th}) \mathcal{D}[a]\rho 
+ \kappa n_\mathrm{th} \mathcal{D}[a^\dagger]\rho
+ \gamma (1+n_\mathrm{th,atom}) \mathcal{D}[\sigma^-]\rho
+ \gamma n_\mathrm{th,atom} \mathcal{D}[\sigma^+]\rho,
\]
with $\mathcal{D}[L]\rho = L \rho L^\dagger - \frac{1}{2} \{ L^\dagger L, \rho\}$.
In the restricted Hilbert space manifold and projecting the dissipative terms following the method of Ref. \cite{DJC2}, the populations evolve as
\begin{align}
\dot{\rho}_{1,1} &= -i g (\rho_{2,1} - \rho_{1,2}) 
- \gamma (1+n_\mathrm{th,atom}) \rho_{1,1} + \gamma n_\mathrm{th,atom} \rho_{3,3}, \nonumber \\
\dot{\rho}_{2,2} &= +i g (\rho_{2,1} - \rho_{1,2}) 
- \kappa (1+n_\mathrm{th}) \rho_{2,2} + \kappa n_\mathrm{th} \rho_{3,3}, \nonumber \\
\dot{\rho}_{3,3} &= \gamma (1+n_\mathrm{th,atom}) \rho_{1,1} + \kappa (1+n_\mathrm{th}) \rho_{2,2} 
- \left(\gamma n_\mathrm{th,atom} + \kappa n_\mathrm{th}\right) \rho_{3,3}, \nonumber
\end{align}
with $\rho_{1,1} + \rho_{2,2} + \rho_{3,3}=1$.
The coherences between atom and cavity mode satisfy
\[
\dot{\rho}_{1,2} = -i \Delta \rho_{1,2} - i g (\rho_{2,2}-\rho_{1,1})
- \frac{\gamma}{2}(1+2 n_\mathrm{th,atom}) \rho_{1,2} - \frac{\kappa}{2}(1+n_\mathrm{th}) \rho_{1,2},
\]
while coherences involving the vacuum state evolve as
\begin{align}
\dot{\rho}_{1,3} &= -i \frac{\Delta}{2} \rho_{1,3} - i g \rho_{2,3} 
- \frac{\gamma}{2} (1+n_\mathrm{th,atom}) \rho_{1,3} - \frac{\kappa}{2} n_\mathrm{th} \rho_{1,3}, \nonumber \\
\dot{\rho}_{2,3} &= +i \frac{\Delta}{2} \rho_{2,3} - i g \rho_{1,3} 
- \frac{\kappa}{2} (1+2 n_\mathrm{th}) \rho_{2,3} - \frac{\gamma}{2} n_\mathrm{th,atom} \rho_{2,3}. \nonumber
\end{align}
The atomic excitation probability and mean photon number are
\[
P_e(t) = \rho_{1,1}(t), \qquad n_\mathrm{ph}(t) \approx \rho_{2,2}(t).
\]

\begin{figure}[t]
\centering
\includegraphics[width=0.8\textwidth]{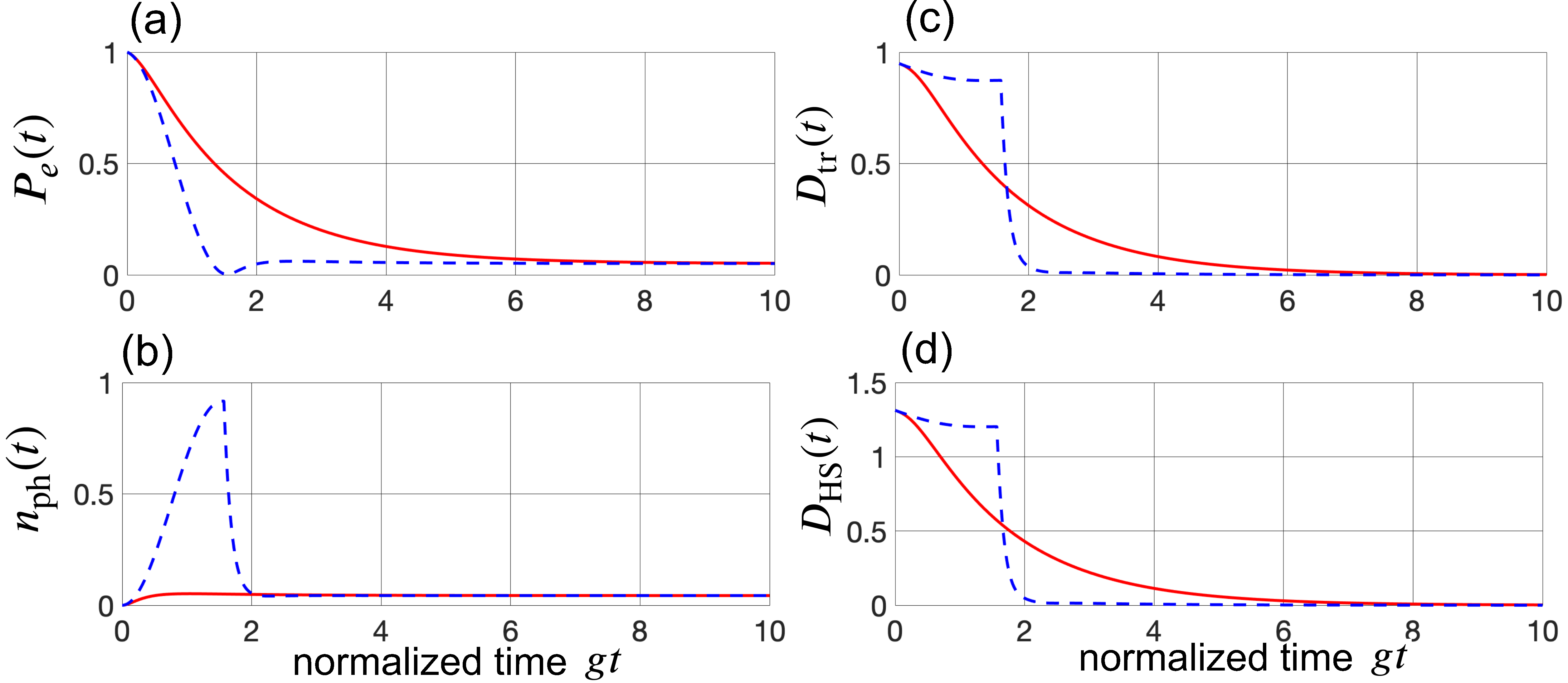}
\caption{
Relaxation dynamics in the dissipative Jaynes--Cummings system with finite temperature baths illustrating the persistence of the quantum Pontus--Mpemba effect. Dashed blue lines correspond to the two-step protocol, where the cavity decay rate is initially low ($\kappa_1 \ll g$) and switched to high loss ($\kappa \gg g$) at time $\tau = \tau_0= \pi/(2 g)$. Solid red lines correspond to the single-step high-loss evolution. 
(a) Excited atomic population $P_e(t)$.  
(b) Cavity photon number $n_{ph}(t)$.  
(c,d) Distance from equilibrium (either trace distance $D_\mathrm{tr}$ or Hilbert--Schmidt distance $D_\mathrm{HS}$).  
The two-step protocol leads to accelerated relaxation: the atomic population decays faster and the system reaches equilibrium earlier than in the single-step high-loss evolution. Parameter values used in the numerical simulations are $\kappa/g=8$, $\kappa_1 /g=0$, $\gamma /g=0.1$ and $\Delta=0$, with mean thermal excitation numbers $n_{{\rm th}}=0.05$ for the photon field and $n_{{\rm th, \; atom}}=0.1$ for the two-level atom.}
\end{figure}
The stationary state $\rho^{(E)}$ can be computed numerically and now includes small populations of both the atomic excited state and the single-photon state. Since the linear dynamical system above differs from the zero-temperature
($n_\mathrm{th}=n_\mathrm{th,atom}=0$) case only through small perturbative
corrections in the thermal excitation numbers, both the stationary state and
the full spectrum of relaxation eigenvalues and eigenvectors are modified only
weakly. Consequently, the Pontus--Mpemba effect is expected to persist for
small but non-vanishing thermal photon and atomic excitation, with only minor
reshaping of the relaxation pathways. This expectation is fully confirmed by
direct numerical simulations of the complete dynamics. As an example, Fig.7, shows the 
dynamical evolution in the dissipative Jaynes--Cummings system with non-zero temperature baths with the same parameter values as in Fig.3, clearly demonstrating the persistence of the quantum Pontus--Mpemba effect. 

}

\end{document}